\documentclass[12pt,a4paper]{article}

\usepackage[dvips]{epsfig}
\DeclareGraphicsExtensions{.eps,.ps}

\begin{document}

\vspace{1cm}
\begin{center}

{\Large{\bf Modelling Food Webs}}

\vspace{0.5cm}

{\em B. Drossel$^1$ and A. J. McKane$^{2,3}$} \\
\bigskip
$^1$Institut f\"ur Festk\"orperphysik, TU Darmstadt, Hochschulstr. 6 \\
D-64289 Darmstadt, Germany \\
$^2$Departments of Physics and Biology, University of Virginia \\
Charlottesville, VA 22904, USA \\
$^3$Department of Theoretical Physics, University of Manchester \\
Manchester M13 9PL, UK\\
\end{center}

\section*{Abstract}

We review theoretical approaches to the understanding of food webs. After an 
overview of the available food web data, we discuss three different classes 
of models. The first class comprise static models, which assign links between 
species according to some simple rule. The second class are dynamical models,
which include the population dynamics of several interacting species. We 
focus on the question of the stability of such webs. The third class are 
species assembly models and evolutionary models, which build webs starting 
from a few species by adding new species through a process of ``invasion'' 
(assembly models) or ``speciation'' (evolutionary models). Evolutionary models 
are found to be capable of building large stable webs. 

\section{Introduction}
\label{intro}

Ecological systems are extremely complex networks, consisting of many
biological species that interact in many different ways, such as mutualism,
competition, parasitism and predator-prey relationships. They have been built
up over long, evolutionary time scales, and in some cases will 
contain extremely ancient structures which hold information on the nature 
of the evolutionary changes which occurred in the distant past. 
Understanding and modelling such complex networks is one of the major
challenges in present-day natural sciences. 

Much research focuses on only a small number of species and their
interactions, such as hosts and their parasites, or the relationship of a 
particular species with its prey or predators. Another important direction 
of research consists in studying larger networks of species by concentrating 
on their feeding relationships and on competition between predators, 
neglecting other types of interaction. Such networks, called food webs, are 
the subject of this review article. We will only be concerned with community 
food webs, which describe these interactions between species in a particular 
habitat, and will not discuss sink webs (species identified when tracing 
interactions down from a particular chosen species) or source webs (species 
identified when tracing interactions up from a particular chosen species). 
From now on community food webs will simply be referred to as food webs, 
or webs.

\begin{figure}
\hspace{3.5cm}\includegraphics*[width=8cm]{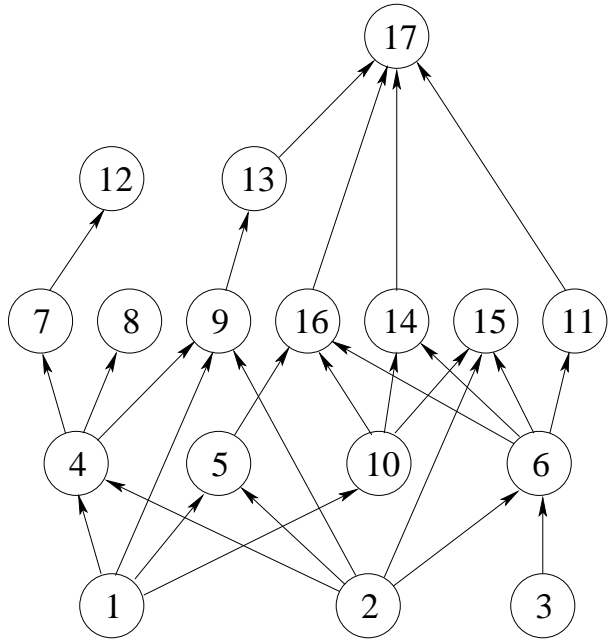}
\caption{Narragansett Bay food web. 1=flagellates, diatoms; 
2=particulate detritus; 3=macroalgae, eelgrass; 
4={\it Acartia}, other copepods; 5=sponges, clams; 6=benthic macrofauna; 
7=ctenophores; 8=meroplankton, fish larvae; 9=pacific menhaden; 
10=bivalves; 11=crabs, lobsters; 12=butterfish; 
13=striped bass, bluefish, mackerel; 14=demersal species; 15=starfish; 
16=flounder; 17=man. (After Kremer and Nixon 1978. 
\emph{A coastal marine ecosystem}, Springer-Verlag, Berlin.) \label{fig1}}
\end{figure}

The early studies of the natural history of a given habitat were
descriptive.  It was not until the third quarter of the nineteenth
century that the idea of listing the basic information on ``what eats
what'' in a particular habitat, and presenting it in the form of a
matrix was born. The usual format has the rows representing predators
and the columns representing prey. The matrix elements might be
numbers which specify the amount of food consumed, or, since such
detail is very rarely known, simply 0's and 1's specifying the
presence or absence of a predator-prey link. The diagrammatic
representation of food webs was introduced some time later (see
Fig.~\ref{fig1} for an example). They consist of vertices representing
species in the food web, with a directed link --- that is a line with
an arrow attached --- from vertex $A$ to vertex $B$, if species $A$ is
eaten by species $B$.
Notice that the direction of the arrows signifies the flow of
resources. These networks illustrate the bare bones of the
predator-prey relationships between the species: they miss much of 
the fine detail such as temporal variations in diet (daily or seasonal), 
but allow the longer term picture of predator-prey relationships to 
emerge. Community food webs cannot include all species in a
habitat (such as all the bacteria living within plants and animals),
but rather focus on a set of different types of species, which are
chosen prior to analyzing their predator-prey relationships.  This
reduction of the rich, distinctive and complex nature of individual
ecosystems tends to be less enthusiastically embraced by field
ecologists, than by theoretical ecologists. For many of those in the
field, the attraction of the study of natural communities is in their
details and unique features. In fact, even when food webs were
published by early investigators, they frequently seemed designed to
illustrate their complexity, rather than to encourage the theoretical
understanding of their form. It was probably this idea of the
distinctiveness and natural complexity of food webs that ensured that
few theoretical investigations were carried out during the century
after the first food webs were constructed.

The change in emphasis in the study of food webs dates from the late 1970's and
early 1980's, when many of the published webs were collected together and
various regularities were noticed for the first time. During roughly the same
period simple models of food webs were formulated. The first models were
dynamic, using simple population dynamics, but without incorporating the 
evolved nature of the web. Static models were also introduced in which species
were simply represented as vertices in a graph and directed links between them
were drawn according to some rule. The structure of the resulting webs could 
then be compared to that of real food webs. 

Obtaining data on dynamic properties is much more difficult than the already
hard task of collecting information on static webs. In any case, many of the
time scales of interest to us will be so long as to be inaccessible by direct
methods. One aspect which is in some sense intermediate between the static and
dynamic descriptions, and which has attracted a large amount of attention, is
the question of the stability of food webs. This is frequently couched in terms
of the complexity of the food web (here meaning a larger web or one with
greater connectance). In other words: does stability increase with web
complexity?

To ecologists working in the 1950's and 1960's the answer was clearly ``yes''.
They pointed to the susceptibility of certain cultivated, or other species 
poor, communities to large scale invasions by pests, and the relative rarity 
of such outbreaks in naturally rich ecosystems as evidence \cite{odu53,elt58}.
Theoretically it was noted that increases in species number or connectance
led rapidly to increases in the available number of food chains or pathways 
in food webs. It then seemed quite compelling to argue that the disruption 
or elimination of only a few of these pathways, if they were numerous, was 
not likely to lead to a complete collapse of the web \cite{mca55}. 

This consensus was disrupted in the early 1970's by investigations into the
linear stability of model ecosystems having random interactions
\cite{gar70}-\cite{may74}. They showed that these model ecosystems became less
stable as species number or connectivity increased. There are two obvious
objections to these conclusions. Firstly, real food webs are not random; they
are highly evolved structures. Secondly, the criterion of linear stability
analysis as applied to population dynamics equations may not be particularly
relevant to real ecosystems, not least because they may not be sufficiently
close to equilibrium for a local condition such as this to apply. 

Over the next few years both objections relating to the validity of the 
randomness assumption and to the use of linear stability analysis for real 
webs, were addressed. Dynamical models with Lotka-Volterra type or more
complicated population dynamics were introduced, and the stability and
dynamical properties of small networks consisting of a few species were
investigated, demonstrating a variety of instances where more complex model
systems were not less stable than simpler ones. Furthermore, models were
introduced that built up a community by a sequence of invasions or
speciations. The most recent of these models lead to the formation of large
stable webs, thus demonstrating explicitly that large complex networks can be
stable. 

The different types of models mentioned above will be reviewed in the following
sections, with an emphasis on more recent work. The earlier phase of
theoretical work has been well documented in a number of books and articles
\cite{pim82}-\cite{coh90}. Before concentrating on theoretical developments, 
however, we will begin (in section \ref{realwebs}) by introducing the basic
concepts used to describe food webs by reference to real webs, and also briefly
allude to the problems in collecting data to construct real webs. 
In section \ref{staticmodels}, we will briefly discuss static food web
models. Section \ref{dynamicmodels} describes the various types of dynamical
models and their contribution to the complexity-stability debate. In section
\ref{evomodels}, we give an overview of models that build food webs by  a
sequence of species invasions or speciations. We include simple toy models as
well as species assembly models and more sophisticated evolutionary models. 
Section \ref{conclusions} concludes the article with a brief overview and a
look to the future. 

\section{Basic properties of food webs}
\label{realwebs}
  
In this section we will characterize food webs by introducing the basic 
features associated with them, quantifying them as much as possible. This 
will allow comparison between real webs and model webs. As the concepts are 
defined, we will also briefly discuss additional aspects such as various 
assumptions made in the definitions, problems or ambiguities associated with 
the definitions, typical values found in real webs or difficulties in obtaining
accurate measurements in the field.

The most primitive concept is the size of the food web, defined by the number 
of species in the web, $S$. In published versions of real webs (see, for 
instance, \cite{coh90}), the terms ``species'' may refer to ``trophic 
species'', which is a collective term for all the species having a common 
set of predators and prey. For this reason general terms such as ``ants'' or 
``algae'' appear; conversely the same species at different stages in its 
life-cycle may belong to a different trophic species. The existence of terms 
such as ``detritus'' or ``dead organic material'' in many webs, is also an
illustration of the difficulty in deciding what to include and what to omit 
from the web. Many earlier studies were not very extensive, and frequently 
the published webs were rather small. For example, the 113 webs listed by
Cohen {\it et al} in their 1990 review \cite{coh90} vary in size from 5 to 48,
with a mean of 17. Since 1990 larger webs have been reported, with several 
containing more than 100 species \cite{mar91}-\cite{mem00}.

The next most important quantity associated with a food web is a
measure of the number of the interactions between the species. This is
frequently taken to be the ratio $L/S$, the total number of
predator-prey links, $L$, divided by the total number of species,
$S$. This property, called the linkage density, seems a fairly natural
choice, but was also favored by some because analysis of the pre-1990
webs suggested that $L/S$ was independent of $S$ and therefore that
this ratio was ``scale invariant''. The value of the slope for the
best fit to $L$ versus $S$ was found to be $1.99 \pm 0.07$, although
there appeared to be a slight tendency for the more recently collected
webs in that set to have a higher value \cite{coh90}. The finding that
the more recently collected webs tend to be larger, suggests the
possibility that the linkage density is not in fact constant, but
slightly increasing with $S$, and that earlier webs were too small to
show this clearly. In fact, this was already noted at the time
\cite{sch89,coh90,pim91}; the scaling relation $L/S \sim S^{\epsilon}$
with $\epsilon$ equal to 0.3 or 0.4 was not ruled out, especially when
the sample included larger webs.

An alternative measure of the number of interactions between species, is the 
connectance, $C$, of a food web defined as the total number of links in the
web divided by the total number of possible links in a web of the same size. 
Since, excluding links from a species to itself (cannibalism), there are
$S(S-1)/2$ possible links in a web of $S$ species, $C = 2L/S(S-1)$. This
quantity was originally introduced by theorists \cite{gar70}-\cite{may74}, 
since it is equal to the probability of finding a non-zero entry in the 
community matrix. Clearly, for large $S$, the power-law scaling mentioned 
above gives $C \sim S^{-1+\epsilon}$, with the scale invariance hypothesis 
leading a hyperbolic $S$-$C$ relation.

During the 1990's the scale invariance hypothesis became more and more 
untenable. The values of the linkage density for the larger post-1990 webs 
ranged from 3.5 to 11.0, compared with the value of 2.0 found for the 
pre-1990 webs \cite{hal93}. The scaling form $L/S \sim S^{\epsilon}$, with a 
value of $\epsilon$ close to or equal to 0.3 or 0.4 suggested above was still
found to be consistent with data \cite{hav92}, although it was also suggested 
\cite{mar92,mar92a} that $\epsilon$ might be as large as 1, leading to
the conclusion that the connectance was independent of the web size. There 
are several reasons why the analysis of more recently collected data might
differ from the earlier results \cite{war94}, but an obvious one involves 
questions of resolution: the earlier webs might be smaller in part because 
many species and/or links were omitted due to incomplete or biased recording.
We will return to this point later in this section, since it is a criticism
which may be applied to other measured quantities. Recent work has confirmed
that the scale invariance hypothesis for the linkage density is not correct, 
but no consensus has emerged to replace it. It may simply be that food webs 
from diverse communities have different characteristics \cite{mur97}, or that 
while this may be the case, much of the disagreement between data from early 
webs and those collected more recently can be explained by the fact that the 
linkage density is very sensitive to sampling effort \cite{ber99}, or that the 
observed patterns can be explained, but not with simple conjectures as scale 
invariance or constant connectance \cite{mon01}.

In addition to quantities which describe the structure of the network,
such as $S$ and $C$, it would be useful to have some characterization
of the type of species in the web.  In order to be meaningful, this
should not be so detailed that generic patterns are not seen, but not
so coarse that it contains little information. The simplest and most
widely used classification is to divide species in the web up into
top, intermediate and basal species.  Top species have no predators,
basal species have no prey --- they obtain all of their resources
directly from the environment --- and intermediate species have both
predators and prey. It is now possible to classify all links in the
web into four classes: links between top species and intermediate
species, top species and basal species, intermediate species and basal
species, and links between intermediate species and other intermediate
species. This gives 7 quantities which contain information about the
biological aspects of the food-web: the proportions of the species
which are top, intermediate and basal (denoted by $T, I$ and $B$) and
the proportion of links between these three types (denoted by, for
example, $IB$, for the proportion of links that go between
intermediate and basal species). Only 5 of these are independent,
because $T+I+B=1$ and also the proportion of the links in all four
classes add up to unity.

This seems to embody just the right amount of information to allow useful
comparison of model webs with real ones. The classification scheme was given
further credence when an analysis of the pre-1990 webs suggested that, like
$L/S$, the proportions $T, I$ and $B$ were independent of web size, having 
values of 29\%, 52\% and 19\% respectively \cite{coh90}. With $L/S$ and
three categories all being independent of web size, it was perhaps not too
surprising that the four types of links, $TI$, $TB$, $IB$ and $II$, were
also found to be the same in webs of different sizes. The reported values were
35\%, 8\%, 27\% and 30\% respectively \cite{coh90}, although, while it is 
true that the data showed no evidence of increasing or decreasing trends, 
the scatter of points looked so random that the conceptual jump to the 
scale-invariance conjecture seemed to be a large one.

As an alternative to the $(T, I, B)$ classification scheme, the proportion of 
the species which are prey, $H$, and the proportions which are predators, $P$, 
may be used. They are related to the previous set by $H=I+B$ and $P=T+I$. 
Recall that only two of the set $\{ T, I, B \}$ are independent and that
$H+P=I+1$ is greater than unity, since intermediate species are both predators
and prey. A frequently quoted statistic for food-webs is the so-called 
predator-prey ratio (in fact, the prey-predator ratio) $H/P$. This seems
to be the property which shows the least change between the earlier, smaller
webs, where it had a value of 0.9, and the more recent, larger webs where it
has a mean value only slightly larger than this \cite{hal93}. There is,
however, a considerable spread in actual values for different webs.

The review of food web patterns by Pimm {\it et. al.} in 1991 \cite{pim91}
was still able to hold on to the belief that many web properties were 
size-independent, though with high variances and with the possible exception 
of the linkage density. However, by the time of the next major review in 1993
\cite{hal93}, it was accepted that $T, I, B$ and the links between them did, 
like $L/S$ or $C$, vary with the size of the web. This change of view was 
mainly prompted by two studies \cite{mar91,hal91}, which took large food webs 
and reduced their resolution by lumping more and more of the species together.
It was found that many of the quantities discussed above were sensitive to 
this aggregation. Although the aggregation criteria which were employed were 
not identical, and thus some of the details of the findings differed, it was 
clear that generally speaking food-webs properties changed with $S$. In 
particular, Martinez \cite{mar91} found that the highly resolved Little Rock 
Lake web had larger $I$, but smaller $T$ and $B$ than the aggregated version. 
The fact that the latter had similar properties to the pre-1990 webs, led him 
to speculate that these webs might too be aggregated versions of larger webs.

These studies were followed up by a re-analysis \cite{mar93} of an earlier
attempt at aggregation \cite{sug89}, which had shown little change with
aggregation, and by a re-analysis \cite{mar94} of the pre-1990 webs
\cite{coh90}. The conclusion of these studies was that aggregated webs and the
earlier, smaller webs all have lower $L/S, I$ and $II$ and higher $T, B$ and
$TB$ than highly resolved or more recently collected, larger webs. The
quantities $TI$ and $IB$ do not seem to change in a consistent way with $S$.
These results can be understood to some degree by beginning with two
observations. Firstly, it is now believed that species without any predator are
very rare, and perhaps non-existent \cite{mar91,pol91}, and thus $T$ will be
tiny in well resolved webs. Secondly, basal species are already rather coarsely
specified, and it would be difficult to aggregate them further. Thus it might
be expected that the proportion of basal species would increase as $S$
decreased. If both $T$ and $B$ decrease as $S$ increases, we would also expect
$TB$ to decrease, and $I$ and $II$ to increase. Presumably $TI$ and $IB$ do not
change in a consistent direction because, unlike $TB$ and $II$, they link types
of species whose proportions change in opposite directions.  For a similar
reason, the predator-prey ratio $(I+B)/(T+I)$ seems to be extremely robust
under aggregation \cite{hal93}.

These ideas can be extrapolated to the limit by considering a food web with 
only two species on the one hand, and the entire global ecosystem on the
other \cite{mar95}. While this might be of dubious validity, the trends
displayed are suggestive. If the two species web consists of a predator and
prey, then $T = 50\%, I = 0\%$ and $B = 50\%$. If we assume that the global
ecosystem has no, or very few, top species, that animals are intermediate 
species and plants basal species, and that animals comprise 95\% of the 
species, then $T = 0\%, I = 95\%$ and $B = 5\%$. These asymptotic values, 
taken together with the previous web results, show a consistent trend of $T$ 
and $B$ decreasing, and $I$ increasing, with $S$, and the possibility that 
food web properties become scale invariant for $S$ larger than about 1000
\cite{mar95}. Studies have also been carried out to investigate other types 
of sampling effects. For example, the threshold for the inclusion of links 
can be varied \cite{win90,mar91} and species can also be omitted (as opposed 
to being aggregated) \cite{gol97}. Once again, it was found that the poorly 
sampled versions of these webs were much more similar to the pre-1990 webs, 
than were the full versions. Conclusions such as these have convinced 
ecologists of the need to be more systematic and methodological in the 
collection of food webs \cite{coh93}-\cite{mar99}.

What are the other food web attributes which field ecologists should be 
looking for? In their review, Cohen {\it et. al.} \cite{coh90} listed five
``laws'' of food webs. Three of these dealt with scale-invariance (of the
linkage density, of $T, I, B$, and of the links between these three types). 
The fourth was that ``food chains are short''. A chain in a food web is the
set of links along a particular path starting from a basal species and ending 
at a top species. The number of links along this path is the length of that
particular food chain. By averaging this over all the chains in a web, a
mean chain length may be assigned to each web. For the food webs listed
in \cite{coh90}, the mean chain length over all 113 webs is 2.88 and the
median of the maximum chain length in each web is 4 links. The observation 
that food chains are short is not new; it is one of the earliest inherent 
tendencies noted in the study of food webs \cite{elt27}. The classic 
explanation is that energy is transmitted very inefficiently up the chain
and after dissipation at more than three or four vertices, is not sufficient 
to sustain predators at the top of the chain \cite{hut59}. This should mean
that productive ecosystems should have longer food chains, but the evidence
for this is mixed \cite{law89}. Other hypotheses are discussed by Pimm
\cite{pim82}. As for the properties discussed earlier, the nature of food
chains in the more recent, larger webs, differ from those appearing in 
Cohen {\it et. al.} \cite{coh90}, being typically much longer. This raises 
the possibility that mean chain lengths are a function of the size of webs. 
Certainly, food chain length decreases when webs are aggregated \cite{hal93}.

The fifth ``law'' was that, excluding cannibalism, cycles are rare 
\cite{coh90}. Cycles are sets of links which end with the same species as they
started from. Of the 113 webs, only 3 contained cycles, and in each case only
one cycle of length 2. By contrast the large Little Rock Lake web \cite{mar91}
contained many cycles. If cycles are at all numerous, the definition of the
length of food chains given above has to be modified, since it is clearly 
ambiguous. Two different algorithms were used to calculate food chain length 
in the case of the Little Rock Lake web. Even when cycle-forming species are 
excluded, reducing the mean chain length, the mean of 7 links is still much 
greater than the 2.88 links for the pre-1990 webs.

A term used frequently in ecology is the ``level'', or ``trophic level'' on 
which a species appears in the food web. This is clearly a useful 
descriptive term, and when it refers to a single food chain it is obviously 
unambiguous: it has a value which is one more than the chain length, that is, 
the number of linkages between it and the basal species in the web 
\cite{pim91}. Equally obvious is the fact that it is not a uniquely defined 
quantity in a web --- there will typically be several routes from the species 
under consideration to basal species. One definition in this case is to list 
all the possible routes and assign the most common (modal) as the trophic 
level \cite{pim82}. Another definition, which seems to us somewhat superior, is
to assign the shortest of the possible routes as the trophic level. This 
choice is based on energy considerations: given the inefficiency of energy
transfer along a chain mentioned above, the most important links are likely 
to be the shortest. This latter definition also has the advantage of being 
unique; in the former case there may be more than one modal value. While the
term ``trophic level'' is used extensively in a qualitative way in studies of
food webs, little is known quantitatively about the number or other attributes
of species on different trophic levels, perhaps because of the absence of a 
single agreed definition. This is unfortunate, because a quantity such as
the fraction of the species on a particular level, is relatively easy to 
obtain from the data and is another attribute which can be compared to models. 
It is also slightly less coarse than the $T, I, B$ designations, and also 
probes the food web hierarchy in a slightly different manner; using the latter
definition above, basal species are always on level 1, but intermediate 
species may also be on level 1, and top species need not be on the highest 
level.

One important property of food webs which rests on the definition of trophic
levels is the degree of omnivory. An omnivorous species is one that feeds on 
more than one trophic level \cite{pim82}. Thus, for instance, a species which 
feeds on its prey's prey is omnivorous. One of the earliest results was that 
omnivory was less common in some types of real webs, than in randomly 
generated webs \cite{pim82}. About 27\% of the species in the pre-1990 webs
are omnivores, but the overall picture is quite confused, in part because
of the different ways that degree of omnivory can be defined \cite{hal93}.
If the assignments of trophic levels has been agreed upon, the most
straightforward index of omnivory is simply the proportion of species which 
are omnivores. Another measure was given by Goldwasser and Roughgarden 
\cite{gol93}. They first determined the statistical distribution of the 
number of links along all pathways from a particular species to the basal 
species. The mean of this distribution, which they termed the trophic height, 
gave a generalized trophic level. The standard deviation, on the other hand, 
gave an indication of the extent to which the species ate on variety of 
different levels, and was used by them as a second index of onmivory.

In any case, omnivory seems to be less common towards the base of a community 
web, and therefore the degree to which sampling favors a particular group
of predators will have a marked effect on the percentage of species which are 
omnivores \cite{hal93}. Some webs have been reported to have a high degree
of omnivory ({\it e.g.} 78\% in \cite{pol91}), so it is again tempting to list 
omnivory as another attribute which has been underestimated in the older web
data, and in fact it has been found to be sensitive to sampling effort 
\cite{gol97}. However, Warren \cite{war94} points out that connectance may be 
a key parameter on which other web characteristics depend, and thus
increase in omnivory may not be independent of the increase in
connectance. Incidently, this type of reasoning may be used to argue
that more highly connected webs may have a higher proportion of
intermediate species (a species is more likely to have links both to
it and from it), more cycles, longer chain length and so on.

The description of food webs given so far in this section has focussed on
static, structural properties of webs. In reality, food webs are dynamical
systems, and links, population sizes, and species composition change with time.
This brings additional difficulties into the quantitative description of web
structure. Empirical data are collected over a certain time which may vary. 
If, for instance, a predator feeds on a certain prey only during harsh 
seasons when other food is scarce, the link to that prey is present only 
temporarily, and only when links to other prey species are absent. Large 
food webs, the data for which have been collected over a long time, may 
therefore overestimate the number of links that are present at a given 
moment in time.

There are, of course, a wealth of field observations of the dynamical behavior
of food webs, but it has not yet been possible to formulate a quantitative,
mathematical description that is generally valid across food webs. There are a
variety of different population dynamics equations containing different
interaction terms, which will be discussed in section \ref{dynamicmodels}. The
discussion about which mathematical form is more appropriate is lively and
diverse. 

This section has not been designed to be an exhaustive review of food webs, 
but rather a summary of the key ideas, concentrating on those that are the 
most relevant for the modelling of webs. The rest of the article will be
devoted to a discussion of the various models of food webs that have been 
put forward. 

\section{Static Models}
\label{staticmodels}

This section describes models that build food webs by assigning links
between species according to some rule, and then evaluate the
properties of the resulting webs. Species are simply represented as
points in space or on a line.

The first such models were modified versions of the random graphs
introduced by Erd\"os and R\'enyi \cite{erd60}, where links are
assigned to randomly chosen pairs of points. Cohen
\cite{coh78,coh90} suggested several models for randomly
generated webs where links have an orientation indicating which of
the two species connected by the link is food for the other one. Links
have no orientation in conventional random graphs. Many properties of
such directed random graphs can be derived analytically, such as the
fraction of top and basal species and the numbers of cycles. The
agreement with data from real webs is not very good.  This is not
surprising, since this simple model has many unrealistic features,
such as the assumption that every species can in principle be the
predator of every other species.

A model that takes into account the fact that some species are higher
up in the food chain than others, has become known under the name
``cascade model'' \cite{coh85a,coh90}. In this model, species are
assigned numbers from 1 to $S$. Each species can prey only on species 
that have a lower number, and it preys on any of these species with a 
probability $d/S$. Here $d$, the density of links per species, is a 
constant which, along with $S$, is the only parameter of the model which
has to be fitted to data. One can easily show that the expected number 
of links in such a food web is $d(S-1)/2$. Therefore, for not too small 
values of $S$, the cascade model predicts that the mean number of species 
should grow linearly with $S$: $L \sim dS/2$. As discussed in section 
\ref{realwebs}, this is consistent with the pre-1990 webs collected in
\cite{coh90}, and with the choice $d=4$ the mean number of links per species 
agrees with the then accepted empirical value close to 2. Other properties, 
such as the fraction of top and basal species, can also be calculated, and 
they are not far from empirical data for the older collections of webs
\cite{coh90}. For example, the values of $T, I$ and $B$ for large $S$ 
asymptote to 26\%, 48\% and 26\% respectively when $d=3.72$. The mean
length of the longest chain increases only slowly with the number of 
species $S$, and it is around 4 for $S$ between $10^3$ and $10^5$.
However, the cascade model seems less good at predicting chain length
statistics, than many of the other measures investigated \cite{coh90}.

In the light of all of the comments made in section \ref{realwebs}  
concerning the difference in trends between data collected in the last
decade or so and older data, it is not surprising that the predictions
of the cascade model have been found to be in disagreement with more
recently collected data \cite{mar92,gol93}. Two of the simplest 
predictions of the cascade model, that food webs are acyclic and that
$L \sim S$, are no longer tenable. In an attempt to generalize the 
cascade model to avoid these and other predictions which are not borne 
out, Cohen constructed 13 alternative versions of the model \cite{coh90a}.
However, in all but one case these were inferior to the original cascade
model in predicting general web properties. For example, models which 
assumed that $L/S \sim S^{\epsilon}$ with $\epsilon = 0.35$ made inferior 
predictions to those models which took $\epsilon = 0$. More recent 
studies have also pointed to deficiencies in the cascade model, especially 
when the assumption of the random distribution of links is viewed in terms 
of aggregated webs \cite{sol98}.

Recently, another static model, called the niche model, was introduced
by Williams and Martinez \cite{wil00}. Just as in the cascade model, 
the species in this model are put in order: a ``niche value'' is assigned 
to them by randomly drawing a number from the interval $[0,1]$. In
contrast with the cascade model, the species are now constrained to 
consume all prey within a range of values whose randomly chosen center 
is less than the consumer's niche value. The size of the range is chosen 
according to a beta distribution with parameters such that the desired 
mean number of links per species results. In contrast to the cascade 
model, species with similar niche values often share consumers, and the 
strict cascade hierarchy is partially relaxed by allowing up to half of 
the consumer's range to include species with niche values higher than the
consumer's value. As in the cascade model there are only two empirical 
parameters: the number of species and the linkage density (or the 
connectance). Evaluating 12 different structural properties of the webs 
generated by the niche model and comparing them to real food web data, 
the authors found that the agreement between the model web and real webs 
is in general much better for the niche model than for the cascade model, 
in particular with respect to features such as cycles and species 
similarities.

In spite of the apparent success at reproducing properties of real
food webs for appropriately chosen parameter values, these static
models cannot give a real explanation of the observed web structures. 
The webs constructed by these models do not result from a dynamical
process; links are not assigned according to some biologically
inspired rule, and the models do not contain any population dynamics.
A good agreement with real data is achieved by capturing
some structural features of real webs, but not by incorporating
underlying biological properties.  In particular, the question of web
stability cannot be addressed in these simple models.
The question of web stability will be discussed in the next section,
where dynamical models are considered, and the question how the
structure of webs might follow from evolutionary dynamics combined with
biological principles, will be explored in section \ref{evomodels}. 

\section{Dynamic models}
\label{dynamicmodels}

The models in the last section attempt to describe food webs as static 
objects, which is after all what nearly all of the data collected is
concerned with. However, it seems more rational to study the kinds of static
structures which emerge from biologically reasonable dynamics, rather than 
attempt to characterize the currently observed webs in terms of simple 
properties of graphs. As stressed in the Introduction, more than one 
time-scale will be relevant in food web dynamics; the long time-scale
evolutionary dynamics and the shorter time-scale population dynamics will
be both important. Evolutionary dynamics will be discussed in the next 
section. In this section we will review the population dynamics of 
predator-prey interactions, with greater emphasis on multispecies communities 
than is traditional in this subject, and with a focus on the question of
under which conditions multispecies communities can be stable. 

We will start with a discussion of the two-species model, which is
frequently as far as most textbooks go. Then, we will generalize the 
two-species dynamic equations to an arbitrary number of species. Finally, 
the  stability of such coupled equations for small webs as function of 
the structural properties of the web, and the types of equations used, 
will be discussed in subsection \ref{stability}. 

\subsection{Two-species models}

In a two-species model a predator (or parasite) depends for
subsistence on a single species of prey (or host) and cannot turn to
an alternative food source. We denote the number of predators at time
$t$ by $P(t)$ and the number of prey by $H(t)$ ($H$ can be thought of
as an abbreviation for ``hosts'' or ``herbivores''). In most cases,
these numbers are understood as individuals per unit area, {\it i.e.}, 
the predator and prey densities. Almost all of the models which are
formulated in terms of differential equations are a special case of
what we will call the standard model \cite{mayn74}-\cite{rou79}
\begin{eqnarray}
\frac{dH}{d\,t} & = & \phi(H) - g(H, P)\,P\,, \nonumber \\
\frac{dP}{d\,t} & = & n(H, P)\,P - d_{P}\,P\,.
\label{standard}
\end{eqnarray}
Here $\phi(H)$ is the growth of the prey in the absence of predators,
$g(H, P)$ is the capture rate of prey per predator, $n(H, P)$ is the
rate at which each predator converts captured prey into predator
births and $d_{P}$ is the (constant) rate at which predators die in
the absence of prey. The function $n(H, P)$, called the numerical
response, which describes how the numbers of new predators relate to
the captured prey, is not usually very well known.  Frequently it is
assumed that a constant fraction of the captured prey are used as
resource to produce new predators, that is, $n(H, P) = \lambda g(H,
P)$, where $\lambda$ is a constant called the ecological
efficiency. Early models assumed that the growth rate of an individual
prey in the absence of predators was constant, that is, $\phi(H) =
rH$, but most models now include intra-species competition by taking
$\phi(H)$ to have the logistic form $\phi(H) = r(1 - H/K)H$, where $K$
is the carrying capacity. With these choices, the type of model is
specified solely by the choice of the function $g(H, P)$, called the
functional response.

The first model of predator-prey dynamics put forward having the form 
(\ref{standard}) was the Lotka-Volterra model which had exponential (not 
logistic) growth of the prey ($\phi(H) = rH$) and a linear functional response
$g(H, P) = aH$, so that the capture rate for an individual predator increased 
linearly with the number of prey \cite{pie77}. This model has the unrealistic 
feature of neutral stability: it contains a limit cycle with an amplitude 
which is determined by the initial conditions, rather than by the parameters 
of the model. Imposing a logistic form for $\phi(H)$ cures this, but only by 
eliminating limit cycles entirely \cite{rou79}. During the 1960's the study 
of the standard model (\ref{standard}), with more realistic forms for the 
functional response began. Rosenzweig and MacArthur \cite{ros63,ros69,mayn74} 
developed a graphical method to determine what functional forms for $g$ and 
$\phi$ gave rise to stable fixed points and limit cycles, although the 
analysis was restricted to functions $g$ which only depended on $H$, and not 
on $P$. A broad conclusion was that the most complete range of behaviors were 
seen in (\ref{standard}) if $\phi(H)$ had the logistic form and if $g(H)$ 
saturated at some constant value for large $H$ (the so-called Type II form).

A specific Type II functional form suggested by Holling \cite{hol65} is widely
used in modelling, partly because of its simplicity, but also because it can
be derived in a reasonably convincing way \cite{has78}. The essential idea
is that the period of searching, $T$, should be divided into true searching 
time, $T_{s}$, and a ``handling time'', $T_{h}$, which represents the time 
taken to eat the prey as well as the time taken afterwards to clean, rest and 
digest the food. Use of $T_{s}$, rather than $T$, in the definition of the
functional response, and the assumption of random encounters between predators
and prey, gives the Holling form
\begin{equation}
g(H) = \frac{aH}{1 + bH}\,,
\label{holling}
\end{equation}
where $a$ and $b$ are constants. Beddington \cite{bed75} extended this idea by
having a second type of ``wasted time'' in addition to the handling time, 
namely time wasted when two predators meet. Incorporating this into the 
definition gave a functional response which depended on the number of 
predators:
\begin{equation}
g(H, P) = \frac{aH}{1 + bH + cP}\,,
\label{beddington}
\end{equation}
where $c$ is another constant. Both forms (\ref{holling}) and 
(\ref{beddington}) are widely accepted as reflecting essential features of 
predator-prey interactions. However, this acceptance is not universal, and 
the traditional arguments used to construct them have been criticized 
\cite{ard89}. The basis of the criticism is that the function $g$ appearing in
the population dynamics equations should be the function calculated on
the same time scale as that of the population dynamics, and not that
calculated on the same time scale as the behavioral response. When
viewed from the slow time scale, prey abundance is assumed to appear as
a continuous function. However, when viewed from the fast behavioral
time scale, prey production is no longer continuous but appears as
successive ``bursts''. Between these bursts, the predators consume the
prey (or the fraction of prey available to predation) by some
mechanism (possibly random search). Thus, for a given number of prey,
each predator's share is reduced if more predators are present. This
suggests that the consumption rate should be a function of prey
abundance {\it per capita}, that is,
\begin{equation}
g(H, P) = \Phi (H/P)\,,
\label{ratio-gen}
\end{equation}
a ratio-dependent functional response. The form of the function $\Phi$
can be deduced by looking at two extreme situations. When the prey is
very abundant, predators feed at a constant maximum rate, so that
$\Phi \rightarrow$ constant, for $H \gg P$. On the other hand, if
predators are very abundant they will consume prey at a constant rate,
so that $g(H, P)P = a'H$ in the limit $H/P \rightarrow 0$.  A simple
form which has this structure is
\begin{equation}
g(H, P) = \frac{a' (H/P)}{1 + b' (H/P)} = 
\frac{a' H}{P + b' H}\,. 
\label{ratio_sp}
\end{equation}
Beddington's form and this specific form of the ratio-dependent functional
response may be written as
\begin{equation}
g(H, P) = \frac{H}{\alpha + \beta H + \gamma P}\,,
\label{both}
\end{equation}
where $\alpha, \beta$ and $\gamma$ are constants. The only difference
is that in the ratio-dependent case $\alpha = 0$. Despite these
similarities, there has been a vigorous discussion in the literature
as to the superiority of one form over the other \cite{han91}-\cite{hui97}. 
The essential differences between the two methods of modelling the functional 
response are discussed in a recent review article written by authors from both
camps \cite{abr00}.

\subsection{Generalized dynamical equations}

The generalization of the population dynamics equations (\ref{standard}),
with realistic growth rates $\phi$ and functional responses $g$, to more
than two species is straightforward for a food chain \cite{has91}, or other 
simple webs, such as two chains with a mobile top predator \cite{pos00}, but 
less obvious for a general web. For this reason virtually all investigators, 
starting with May \cite{may72}, who have studied the population dynamics for 
a general web have used Lotka-Volterra dynamics. However the well-known 
unsatisfactory features of these equations \cite{may74}, together with a 
desire for greater realism, have resulted in some suggested versions of 
population dynamics which go beyond the Lotka-Volterra scheme 
\cite{get84}-\cite{drossel01}.

Let us begin with the Lotka-Volterra equations. If $N_i$ is the population 
size or population density of species $i$, the Lotka-Volterra equations for
a general web may be written as
\begin{equation} 
\frac{dN_{i}(t)}{dt} = N_i\left(b_i + \sum_{j} a_{ij}N_{j} \right)\, ,
\label{LV}
\end{equation}
where $b_i$ is a positive growth rate for basal species, and a
negative death rate for the other species. The $b_i$ and the
interaction coefficients $a_{ij}$ are constants, independent of the
population sizes.  There are three different possible contributions to
$\sum_{j} a_{ij}N_{j}$: (i) As mentioned in the context of
two--species models, many authors include a logistic term for the
basal species, implying $a_{ii} < 0$ for basal species, and zero for
the other species. (ii) Some authors using Lotka--Volterra models
include competition between two predators that share the same prey,
{\it i.e.}, $a_{ij}<0$ whenever $i$ and $j$ have a prey in
common. (iii) The most important contributions are the predator--prey
terms. If $i$ is a predator and $j$ is one if its prey species, then
$-a_{ji}N_j$ is the functional response $g_{ij}$, {\it i.e.}, the
number of individuals of species $j$ consumed per unit time by an
individual of species $i$. Often, the identity $-a_{ji}=\lambda
a_{ij}$ is used, but some Lotka--Volterra models have independent
random (positive) numbers for $a_{ij}$ and $-a_{ji}$, and some models
do not even impose opposite signs for $a_{ij}$ and $a_{ji}$. 

We will restrict our discussion of general non-Lotka-Volterra type
equations to those that satisfy the balance equations
\begin{equation}
\frac{dN_{i}(t)}{dt} = \lambda\sum_{j}N_{i}(t)g_{ij}(t) - 
\sum_{j} N_{j}(t)g_{ji}(t) - d_{i}N_{i}(t)\,.  
\label{balance}
\end{equation}
These equations are in many ways the natural generalizations of 
(\ref{standard}), with the first term on the right-hand side representing
the growth in numbers of species $i$ due to predation on other species, the 
second term the decrease in numbers due to predation by other species, and 
the last term the constant rate of death of individuals of species $i$, in 
the absence of interactions with other species. Where there is no 
predator-prey relationship between species $i$ and species $j$, $g_{ij}$ 
is zero. There are two minor variants on (\ref{balance}): the basal 
species may be treated differently from the other species, and given a 
positive growth term to represent feeding off the environment, or the 
environment may be included as a ``species 0'' and these growth terms 
represented by functional responses $g_{i0}$.

Apart from the constant death rates $d_{i}$ and the ecological efficiency,
$\lambda$, the model is completely specified once the functional responses 
have been chosen. Arditi and Michalski \cite{ard96} have pointed out that 
these generalized functional responses, if they are to be logically 
consistent, must leave the balance equations invariant if two identical 
species are aggregated into a single species. The obvious generalized form 
of the Holling type functional response, Eq.~(\ref{holling}), is
\begin{equation}
g_{ij} = \frac{a_{ij} N_j}{1+\sum_k b_{ik}N_k}\,,
\label{generalholling}
\end{equation}
where the sum in the denominator is taken over all prey $k$ of species
$i$.

Generalizations of more complicated functional responses can only be
found in the recent literature.  Arditi and Michalski
\cite{ard96} suggest the following generalized Beddington form:
\begin{equation}
g_{ij} = \frac{a_{ij} N_j}{1+\sum_k b_{ik}N_k+\sum_lc_{il}N_l}\,,
\label{generalbeddington}
\end{equation}
where the first sum is again taken over
 all prey $k$ of species $i$, and the second sum is taken over all
 those predator species $l$ that share a prey with $i$.

A possible generalization of the ratio-dependent functional response
 results from Eq.~(\ref{generalbeddington}) if the 1 in the
 denominator is cancelled. However, as Arditi and Michalski
\cite{ard96} point out, the idea that predators share the prey, which
 led to the introduction of ratio-dependent functional responses, is
better reflected by the following expression,
\begin{equation}
g_{ij} = \frac{a_{ij} N_j^{r(i)}}{N_i+\sum_{k\in R(i)}b_{ik}N_k^{r(i)}}\,,
\label{generalratio}
\end{equation}
with the self-consistent conditions 
$$N_j^{r(i)}=\frac{\beta_{ji}N_i^{C(j)}N_j}{\sum_{k\in
C(j)}\beta_{jk}N_k^{C(j)}},\quad
N_k^{C(j)}=\frac{h_{jk}N_j^{r(k)}N_k}{\sum_{l\in R(k)}h_{lk}N_l^{r(k)}}\,
.$$
Here $\beta_{ij}$ is the efficiency of  predator $i$ at consuming
species $j$, $h_{ij}$ is the relative preference of predator
$i$ for prey $j$, $R(i)$ are the prey species for predator $i$,
$C(i)$ are the species predating on prey $i$, $N_j^{r(i)}$ is the part
of species $j$ that is currently being accessed as resource by species
$i$ and $N_k^{C(j)}$ is the part of species $k$ that is currently
acting as consumer of species $j$. An interesting consequence of this
implicit form of the functional response is that not all the links that
are in principle possible are realized, by a long way. This is a very 
realistic feature of the model, since species typically feed on those 
prey that are most easily available, and resort to other prey only 
during periods of food shortage. Arditi and Michalski \cite{ard96} also 
found that small food webs with this generalized ratio-dependent 
functional response are far less sensitive to the aggregation of species 
than webs with prey-dependent functional responses. 

A shortcoming of model (\ref{generalratio}) is that the predator
preferences $h_{ij}$ are constants that are independent of prey
availability. In reality, one can expect that predators assign more
effort to those prey from which they obtain more food per unit effort,
so that a stationary point is reached only when a predator obtains
from each prey the same amount of food per unit effort. This condition is
implemented in the generalized ratio-dependent functional response
suggested by Drossel, Higgs, and McKane \cite{drossel01}, Eq.~(\ref{gij}),
which is discussed in the next section.

\subsection{The complexity-stability debate}
\label{stability}

So far in this section we have surveyed the kinds of population
dynamics equations which are frequently applied to the modelling of
predator-prey systems. As is usual, we have assumed that the
parameters of the various models are given, but for a large community
these may be hundreds in number.  Obviously some way of specifying the
parameters is required, and it is at this stage that we move into the
question of food-web modelling, since many of these parameters will be
related to the underlying web structure. The methods that have been
used to go beyond pure population dynamics to incorporate food-web
structure fall into three classes (see, for instance,\cite{yod89}, who
defines the first two classes). The first class, which is the object
of this subsection, studies the stability of small webs as function of
their structure, of the choice of dynamic equations, or of the choice
of parameter values. The motivation for this type of study is the
intuition that real food webs must be stable, Part of this program
involves defining exactly what is meant by ``stable''. The second
class, which will be studied in section \ref{assembly}, assembles
communities from a very small original system by bringing in species
from a ``species pool'', and if they can add to the community in a
stable way, they are incorporated into the system. In this way, larger
ecosystem can, in principle, be built up. Third, in evolutionary
models (see section
\ref{realisticmodels}), a community is built up not from a preexisting pool
of species, but by modification (``mutation'') of existing species.

The first attempt to write down mathematical equations for the dynamics of
food webs and to study their stability, is due to May \cite{may72}. 
May performed a linear stability analysis of the 
population sizes around a supposed equilibrium point:
\begin{equation}
\frac{d\ }{dt}\left( \delta N_{i} \right) = \sum_{j}\alpha_{ij}\,\delta N_j \,,
\label{may}
\end{equation}
where $\delta N_i$ is the deviation of the population size of species
$i$ from its equilibrium value and $\alpha_{ij}$ is the community
matrix. In this way he avoided specifying the underlying population
dynamics equations, but was constrained to stay near equilibrium. The
choice of web structure is equivalent to the choice of the
$\alpha_{ij}$. May chose the diagonal elements of the matrix to be
$-1$. The other elements were taken to be zero with probability
$1-h$. With probability $h$, they had a random nonzero value chosen
from a distribution of width $\alpha$, so that $\alpha$ is a measure
of the average interaction strength. Using results from random matrix
theory, he found that ecosystems that are initially stable will become
less stable ({\it i.e.}, the initially negative eigenvalues of the 
community matrix move towards zero) when $\alpha (SC)^{1/2}$ is
increased. Furthermore, Eq.~(\ref{may}) will almost certainly be stable
if $\alpha (SC)^{1/2} < 1$, and almost certainly be unstable if
$\alpha (SC)^{1/2} > 1$.  This finding spurred on much of the interest
in the relationship between $C$ and $S$ discussed earlier. The belief
that webs with high connectance were unstable supplied a reason why
webs with large $C$ were not observed. On the other hand, the result
was hard to reconcile with the increasing evidence for the scaling
relation $C \sim S^{-1+\epsilon},\,\epsilon > 0$ unless $\alpha$ was
very small, there were complaints from field ecologists that the webs
which they had been observing for many years should be unstable
according to the May criterion \cite{pol91}, and there were discussions 
concerning the mathematical basis of the result \cite{has82}-\cite{erd90}.

Although May's work was interesting because it broke new ground, there
were obviously several weak points in the analysis. One was the lack
of biological realism assigned to the web. It was argued that the web
structure should be ``plausible'', and not just randomly generated
\cite{dea75,law78}. It was suggested, for example, that food webs with
``realistic'', rather than random, structures had more chance of being
stable \cite{law78}. These ideas were made more concrete by Yodzis
\cite{yod81}, who constructed ``plausible community matrices'' by
using the topologies of real webs, with the correct sign and an
estimate of the magnitude of the strength of the links. He then showed
that in every case where community matrices were plausible, disrupted
forms of these matrices, which no longer represented real communities,
were less stable. Other authors \cite{rob81,tay88} started with a large
random Lotka-Volterra system (of the order of 50 species) and
successively removed those species that were least stable, until a
stable smaller food web was obtained, which typically had more positive
coefficients than random networks. Still other authors investigated
the stability of small Lotka-Volterra food webs (typically 4 to 10
species) as a function of the connectivity pattern and the link
strengths. For Lotka-Volterra systems, one simply has $\alpha_{ij} =
N_i^* a_{ij}$, where $N_i^*$ is the equilibrium population size of
species $i$. Taking into account differences in body size between
predators and their prey, Pimm and Lawton \cite{pim78} found that webs
with more omnivory ({\it i.e.}, more links) are not always less likely 
to be stable. The exception occurs in webs where a ``predator'' is a 
small parasite $i$ of a large host, $j$, in which case 
$|\alpha_{ij}|$ is much larger than $|\alpha_{ji}|$. De Angelis
\cite{dea75} found that small webs are more stable when the ecological
efficiency $\lambda$ is smaller, when species on higher trophic levels
have strong self-limitation ({\it i.e.,} a strong negative $a_{ii}$), or when
the predator population dynamics have little impact on their prey.

Very recent evidence suggests that models with more realistic
functional responses tend to be more stable than Lotka-Volterra
systems. In those models, the community matrix $\alpha_{ij}$ has
no simple relation to the coefficients in the dynamical equations, and
its values can therefore be expected to be far from random, even if the
parameters in the dynamical equations are chosen in some random way. 
This was demonstrated explicitly by Pelletier \cite{pel00}, who
studied a system of $n$ basal species (prey) and $n$ predator species
feeding on these prey, choosing a  functional response of the form
$$g_{ij}= a_{ij} N_j \frac{na_{ij}N_j}{\sum_ka_{ik}N_k}\, .$$ In this
way, a predator can assign more weight to a prey from which it obtains
more food. Pelletier found that 85\% of these types of food webs
(with random values for $a_{ij}$) are stable, irrespective of the
value of $n$, in contrast to Lotka-Volterra systems, where the
percentage of stable webs decreases quickly with $n$. 
We will see in the next section that an evolutionary model that
uses generalized ratio-dependent functional responses, can build larger
food webs than a Lotka-Volterra type model, indicating again that
Lotka-Volterra systems are less stable than more realistic ones.

Another weak point of May's analysis is the use of linear stability
analysis. Clearly, a model ecosystem need not be at a stable
equilibrium point in order to be realistic, but may instead be on a
limit cycle or even a chaotic trajectory, as long as the fluctuations are
small enough that no species get close to extinction. In such a
situation, the question as to whether more complex ecosystems are 
more stable takes the form ``under what conditions have more complex
systems smaller fluctuations in population sizes''. Intuitive
arguments were put forward that the addition of weak links to an
existing web with a strong predator-prey coupling should have a
dampening effect on the population oscillations of the strongly
coupled predator-prey pair. The reason is that the predator can feed
on an alternative prey to which it has a weak link when its main prey
becomes low in population size, allowing the main prey to
recover. Similarly, a weakly linked alternate predator can increase in
population size when the main predator decreases, thus preventing a
large oscillation in prey population. These arguments are supported by
the numerical study of models for small food webs with several weak
links. Using a Holling-type generalized functional response, McCann,
Hastings, and Huxel \cite{cann98} found that the weak links have
indeed a stabilizing effect on the model dynamics. Polis \cite{pol98}
suggests that the chosen form of the functional response is important
for the result, since it makes it impossible for a predator to
maintain a high feeding efficiency on many prey at the same time (in
contrast to Lotka-Volterra systems). Field data seem to support the
hypothesis that stable food webs have many more weak links than strong
links \cite{cann98}. 

Some authors point out that species rich communities should have less
community-level variability ({\it i.e.},  relative fluctuation of the
combined density of all species sharing the same ecological role)
than species poor communities, where the oscillations of one species
cannot be counterbalanced by different oscillations of another
species.  This concept of community-level stability is supported by
numerical simulation of a Lotka-Volterra type model \cite{ive00}, and
is reviewed by Loreau in \cite{lor00}. 

An alternative definition of stability, called ``species deletion stability'',
which might have more direct relevance to real webs was introduced by Pimm in 
1979 \cite{pim79}. An ecosystem is defined to be species deletion stable if, 
when a species is removed from the web, all the remaining species remain at
a stable equilibrium involving only positive densities. Species deletion 
stability decreases with increasing numbers of species and connectance,
{\it i.e.,} decreases with complexity \cite{pim79}, but it also depends 
crucially on which species are selected for removal \cite{pim80}. A 
quantitative measure of the deletion stability of a web is provided by $S_{d}$,
the fraction of species for which the web is species deletion stable. However,
it should be noted that data on experimental species removal show that many 
real species are not species deletion stable \cite{pim80}. Recent work
\cite{bor00} showed that the risk of additional species deletions,
following the loss of one species in model food webs, decreases with
biodiversity. A review of the relation between the complexity and
stability of an ecosystem \cite{pim84} concluded that much of the
confusion in the literature to date arose because of the different
meanings given to the terms ``complexity'' and ``stability''; many
different definitions of perturbations and persistence are possible,
and only a few are appropriate for real webs. One of the more fruitful
of these has been the idea of ``permanence'' \cite{jan87,law92}, which
will be explored in more detail in the following section, when
assembly models are discussed. Recently the diversity-stability debate 
has been reviewed by McCann \cite{can00}.

\section{Assembly models and evolutionary models}
\label{evomodels}

This section describes models for food webs which incorporate longer time
scales.  In contrast to the models presented so far, they allow for
the ongoing introduction of new species (due to immigration or
speciation) and for species extinctions. As a consequence, the
composition and structure of the web changes with time. Studies of
assembly models and evolutionary models focus mainly on the features
of the food web after a sufficiently long time, when the size of the
food web and other properties cease to change in a systematic manner.

While the static models presented in section
\ref{staticmodels} are only concerned with web structure, but cannot address 
web stability, and while the dynamic models presented in section
\ref{dynamicmodels} focus on the stability of web subunits but do not
deal with the overall web structure, the models presented in this
section combine the two aspects of web structure and web
stability. Another advantage of assembly models and evolutionary
models is that links between species and interaction strengths are
shaped by the web's history, instead of being assigned in an ad-hoc
manner as in the other two types of models.

Assembly models and evolutionary models can be divided into three
classes, which will be presented in the following subsections. The
first class comprises toy models that resemble to some extent the
static models discussed in section
\ref{staticmodels}. They ignore population sizes, and species, and
links are added and removed according to simple rules. The structure
of the resulting webs is usually different from the structure of
real food webs, as described in section \ref{realwebs}. However, the
main focus of these models is on species extinctions rather than on
the food web structure, and these extinction events bear some
similarity to those seen in the fossil record \cite{raup86}.

The other two classes of models are more realistic, as they take
population sizes into account and include such important features as
competition for food and link strength, which are not part of the toy
models.  The second class of models are species assembly models,
which, starting from a small initial system, bring in new species from
a species pool that are incorporated in the system if they add to the
community in a stable way. These species assembly models, which
typically lead to an uninvadable, stable system, will be discussed in
subsection
\ref{assembly}.  The third class of models, reviewed in subsection 
\ref{realisticmodels} are inspired by biological
evolution through modification of existing species. Just as the
assembly models, they start from a small set of species, and then add
new species, which are obtained by modifying existing species.  In
spite of differences in the population dynamics, the different
evolutionary models lead to similar and realistic food web shapes.

\subsection{Toy models}
\label{toymodels}

The purpose of introducing evolutionary toy models was not so much to
reproduce realistic web structures, but rather to study the 
large-scale dynamics of species extinctions.  
Species are usually characterized by a number which is related to
their fitness, and they become extinct when this number falls below a
threshold value. The web structure can be a regular lattice, as in the
Bak--Sneppen model \cite{bak93}, or a fully connected web, as in the 
Sol\'e--Manrubia model \cite{sole96}, or new links are added together
with a new species according to some rule, as in the 
Slanina--Kotrla model \cite{slanina99} and the Amaral-Meyer model
\cite{amaral99}. Fitness changes are triggered  by changes in species linked
to a given species, and they also include a stochastic component.  An
overview of all these models was given by Newman and Palmer
\cite{newman99} and by Drossel
\cite{drossel01a}. 

Since the links in these models are in most cases
not understood to be feeding relations but interactions of any type,
their connection to food webs is only superficial. In the following, we
give a description only of the model by Amaral and Meyer, which is the
one closest to food webs, since it places
the species in trophic layers, with links indicating which species
feeds on which other species. The model is defined as follows: Species
can occupy niches in a model ecosystem with $L$ levels in the food
chain, and $N$ niches in each level.  Species from the first level
$l=0$ do not depend on other species for their food, while species on
the higher levels $l$ each feed on $k$ or less species in the level
$l-1$. Changes in the system occur due to two processes: (i) Creation
of new species with a rate $\mu$ for each existing species. The new
species becomes located at a randomly chosen niche in the same level
or in one of the two neighboring levels of the parent species. If the
new species arises in a level $l>0$, $k$ species are chosen at random
from the layer below as prey. A species never changes its prey after
this initial choice.  (ii) Extinction: At rate $p$, species in the
first level $l=0$ become extinct. Any species in layer $l=1$ and
subsequently in higher levels, for which all preys have become
extinct, also become extinct immediately. This rule leads to
avalanches of extinction that may extend through several
layers. Amaral and Meyer found from computer simulations that the size
distribution of these extinction avalanches is given by a power law
$n(s) \sim s^{-\tau}$ with $\tau \simeq 2$. This result $\tau=2$,
which was confirmed by an analytical calculation by Drossel
\cite{drossel98}, is compatible with the findings of paleontologists
that species extinction events of all sizes have occurred in the
geological past \cite{raup86}. A more detailed study which also
includes the taxonomy generated by the model is given by Camacho and Sol\'e
\cite{camacho00}. 

Large extinction avalanches are also found in the other
toy models mentioned above, and they imply that the internal dynamics
of ecosystems place them at the border of stability such that small
triggers can have large consequences. However, the more widely accepted 
view seems to be that ecosystems in themselves are rather stable, but that
external events like meteorite impacts or changes in the sea level are
to blame for the large extinctions in the geological past. If this is
correct, the simple toy models miss important ingredients that are
present in real ecosystems. The more realistic models described in the
next subsection lead to food webs that are much more stable.

\subsection{Species assembly models}
\label{assembly}

The more realistic assembly and evolutionary models, which will be
discussed in the remainder of this section, include population
dynamics. They have two time-scales, which are assumed to be
separated. On the faster, ecological time scale, population sizes
change until they reach fixed points or stationary orbits. On the
slower time scale, new species are introduced by immigration (assembly 
models) or by modifying one or a few individuals of an existing 
species (evolutionary models). After introduction of the new species, 
the population dynamics may either drive this new species to extinction, 
or the new species becomes established, while possibly one or a few 
other species become extinct. Even if no species become extinct, the 
food web may become rearranged, with species abandoning one prey or 
choosing an additional prey.

Species assembly models take into account that real ecosystems, for
instance on an island, are often built up by species immigration. 
Starting with either one or a few
species, species from a ``species pool'' are added to the system, and
they remain in it if the resulting system is stable. Energetically
constrained community assembly was modelled by Yozdis
\cite{yod81,yod84}. Starting with $N$ basal species each of which has
a ``production'' $P$, new species are introduced one by one. The
required energy intake $e$ of a new species is chosen from a given
probability distribution, and the prey species are chosen one by one
with probabilities proportional to their unused production. If a prey
has a randomly chosen fraction of its production available, this prey
is utilized by the new species, and further links to preys are added
until the energy needs of the new species are satisfied. The assembly
process ends when the total unutilized production falls below a minimum
value. The resulting webs have properties that agree well with real
web data. 

In all the other models, population dynamics is modelled via
Lotka-Volterra equations. The species pool is usually a set of no more
than 25 basal species (``plants''), and the same number of
``herbivores'', ``carnivores'' and top predators, with interaction
coefficients between neighboring layers assigned according to some
random rule (sometimes taking the larger body size of consumers into
consideration or trying to include ``specialists'' that feed on only
one prey as well as ``generalists'' that can feed on several
prey). The third or fourth trophic level (carnivores and top
predators) are missing in some of the models. Although this species
pool is usually interpreted as stemming from a large ecosystem, like
the mainland, no stability criteria or other criteria inspired from
real large webs are applied to it.

After adding a new species with an initially small population size to
the system, one of the three following things can happen: (i) The new
species increases and coexists with all the other species. (ii) The
new species remains in the system, but one or more other species go
extinct. (iii) The new species goes extinct.  Numerical integration of
Lotka-Volterra equations, combined with the criterion of local
stability, were used by Post and Pimm \cite{pos83} and by Drake
\cite{dra88,dra90}, to construct webs of typically less than 20 species.
Since the numerical integration of large Lotka-Volterra systems is
very inefficient \cite{mor96}, other authors \cite{law96,mor97} use
the concept of permanence in order to find the composition of the new
community after species addition. An ecological community is permanent
if all species remain at a positive finite density when the density of
each is started at a positive finite value.  For Lotka-Volterra
systems, the permanence of a system can be quickly tested using only
two criteria.  Clearly, this criterion of permanence is too
strict, since real systems never explore the full space of possible
population sizes, and since it seems implausible that for real
communities even very unusual combinations of population sizes should
not result in species extinctions.

After some time, an invasion-resistant state is achieved, the
properties of which can be evaluated as function of the properties of
the species pool. The invasion-resistant state may be a single
community, or (in a minority of cases) a cyclic sequence of
communities. Typically, the size of the resulting community increases
with the size of the pool, but saturates when the pool size becomes
large \cite{mor97}.

Lockwood {\it et al} \cite{loc97} showed that if subsequent invasions
follow rapidly (instead of waiting for a stable species configuration
after each invasion), before the system can achieve an equilibrium state,
communities do not evolve towards an invasion-resistant state, but
move through complex cycles of composition, where each species gets
its turn. A recent review on community assembly, with the main focus on
phenomenology rather than models, is \cite{bel99}.

To summarize, the species assembly models put forward so far are
capable of generating intermediate-size webs with a predetermined
number of trophic layers. There are several drawbacks of these
models. First, so far only Lotka-Volterra equations have been
used. However, since equations using other functional responses are
known to be more realistic and more stable, it would be worthwhile to
investigate species-assembly models with other functional
responses. Second, the species pool is not very large, thus limiting
the number of possible modifications of the web. It might well be
possible that with a much larger pool, the webs would not evolve
towards an invasion-resistant state. Third, the species pool is
composed of random species. Since the assembled web will consist after
some time of species that are in some respect adapted to each other,
it is very unlikely that a randomly defined additional species could
invade the system. In contrast, real species pools contain species
that have evolved to be able to survive well in the presence of other
species from the pool. A real species pool, even when not large, will
therefore contain many more species that can invade the ecosystem under
consideration, than do the random species pool used in the models. The
evolutionary models presented in the next subsection have no species pool
at all, but they introduce new species as modifications of existing
ones. New species are therefore much more likely to fit into the
existing ecosystem than the randomly generated species in assembly
models.

\subsection{Evolutionary models}
\label{realisticmodels}

Evolutionary food web models introduce new species as variations of
existing ones.  The first evolutionary food web model that includes
population dynamics was introduced by Caldarelli, Higgs and McKane
\cite{caldarelli98}.  Species in this model can be characterized as
binary strings, with each bit representing a feature that is either
present (1) or absent (0) in a species. This representation gives a
measure of similarity between species (the number of features they
have in common) and allows for ``mutations'' by randomly swapping a
1-bit and a 0-bit ({\it i.e.}, by replacing one feature with
another). ``Scores'' between two species are obtained by multiplying
the two feature vectors to the right and left of an asymmetric random
matrix that is chosen at the beginning of the simulation. Positive
scores indicate that the first species can feed on the second species,
and negative scores mean that the first species is eaten by the
second. The external resources are represented as an additional
species of fixed (and large) population size, which does not feed on
any species. The population dynamics are simple: at each time step, a
fixed percentage of every species that has at least one predator is
eaten by the predators of that species.  The prey is divided such that
all those predators that have a score within a certain narrow range of
the maximum score against a prey species obtain a share, the size of
which depends linearly on the score. This means that species do not
feed on all prey species they can potentially feed on. As the web
changes and evolves, the prey species eaten by a given predator can
change.  Since the dynamical equations are linear, they quickly reach
a fixed point. Then, a randomly chosen individual is ``mutated'', and the
new population sizes are calculated. Starting with one species
and the external resources, large webs can be built. After some time,
a stable species configuration is reached such that no ``mutant'' can
become established. The parameters of the model can be chosen such
that the fractions of top and bottom species, the numbers of links per
species and other properties of the webs are very similar to those of
real food webs.

In a subsequent paper by Drossel, Higgs and McKane
\cite{drossel01}, a modified version of this model was introduced,
which contains more realistic population equations.  For all species
equations of the form Eq.~(\ref{balance}) were used. As in the
previous model, $\lambda=0.1$ was chosen, and the external resources
were modelled as an additional species with a large and fixed
population size. Apart from the external resources, all species have a
death rate $d_{i}=1$ and a ratio-dependent functional response of the
form
\begin{equation}
g_{ij}(t) = \frac{S_{ij}f_{ij}(t)N_j(t)}{bN_j(t)
+\sum_k \alpha_{ki}S_{kj}f_{kj}(t)N_k(t)}\,.
\label{gij}
\end{equation}
The $S_{ij}$ are the above--mentioned scores, and  $f_{ij}$
is the fraction of its
effort (or available searching time) that species $i$ puts into preying on
species $j$. These efforts must satisfy $\sum_j f_{ij}=1$ for all $i$,
and they are determined self--consistently from the condition
\begin{equation}
f_{ij}(t) = \frac{g_{ij}(t)}{\sum_k g_{ik}(t)}.\label{eff}
\end{equation}
This condition is such that no individual can increase its energy
intake by putting more effort into a different prey. The parameters 
$\alpha_{ki}$ give a measure of the strength of competition between 
species $k$ and $i$. They are equal to 1 for $i=k$, and a linear function 
of the overlap between species $i$ and $k$ ({\it i.e.}, of the number of 
features that $i$ and $k$ have in common) for $i \neq k$.

\begin{figure}
\hspace{2cm}\includegraphics*[width=12cm]{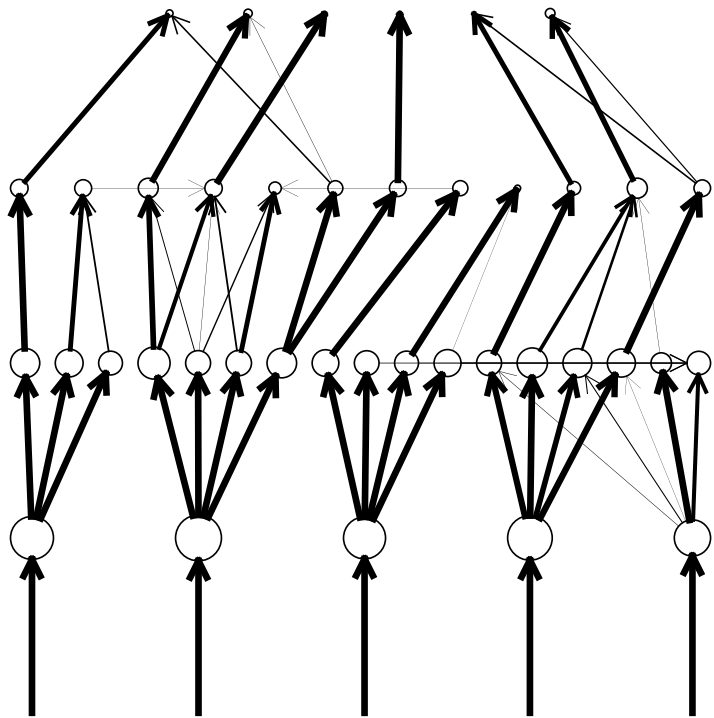}
\caption{Example of a food web generated by the evolutionary model 
\cite{drossel01}. The radius of the circles is proportional to the
logarithm of the population size, and the thickness of a link is a 
measure of the energy flow along that link. \label{fig2}}
\end{figure}

In computer simulations of the model, the population sizes quickly
reach a fixed point. As discussed in the previous section, this is
generally not the case for Lotka--Volterra type population
dynamics. The effects of competition, of predator saturation, the
ratio-dependent functional response, and the ability to assign more
effort to searching for the better prey species, may all play a
crucial role in stabilizing the population dynamics.  Using the same
evolutionary dynamics as in the previous model, large food webs can
again be built that consist of several hundreds of species.  Just as
with the simpler population dynamics equations of the previous model,
the properties of the food webs agree well with the empirical ones. In
contrast to the above model, which has simpler population dynamics, no 
stable species configuration is reached, but there is ongoing species 
creation and extinction, even after a long time. However, no more than 
a few species become extinct at the same time, and the size distribution 
of extinction events has a sharp exponential cutoff. The evolutionary 
dynamics of the model, combined with the population dynamics, thus 
create large stable webs, which have ongoing changes due to species 
overturn, but do not show strong responses to small perturbations. More 
recent studies of this model can be found in \cite{quince01,quince02}.
Figure \ref{fig2} shows an example of a food web generated by this model.

A different dynamical model, which uses Lotka--Volterra type
equations, was introduced by L\"assig {\it et al} \cite{laessig01}. In
contrast to standard Lotka--Volterra equations, where the quadratic
term is only used for predator--prey relationships, these authors also
include a predator--predator competition term. The $a_{ij}$ in
Eq.~(\ref{LV}) are equal to a
constant $\gamma_+$ if $j$ is predator of $i$, and $-\gamma_-$ (with
$0<\gamma_-<\gamma_+$) if $i$ is predator of $j$. If $i$ and $j$ have
a prey in common, then a competition term is added to $a_{ij}$, which
is $-1$ for $i=j$ and $-\beta \rho_{ij}$ otherwise, with $\beta$ being
a constant smaller than 1, and $\rho_{ij}$ the link overlap between
the two species, which is defined as the geometric mean between the
fraction of $i$'s prey species that it shares with $j$ and the
fraction of $j$'s prey species that it shares with $i$. All species
have the same death rate. The external resources are represented as a
few species with a positive growth rate and no prey.

The ``mutations'' which generate the evolutionary dynamics in this
model consist in a change of a predation link for an individual. Using
a mean--field approximation, the authors analyzed this model
analytically and obtained web structures which are very similar to
realistic food webs. Computer simulations of this model are given in
\cite{bastolla02}. However, up to now only simulation results for one
trophic layer of species have been published. All these species feed on the
external resources, which are modelled as permanent species with a
constant growth rate. The computer simulations show a constant
turnover of species, with considerable fluctuations in the total
number of species. This is in strong contrast to model
\cite{drossel01}, where simulations of one trophic layer always
converged to an uninvadable species constellation. This confirms again
the finding that Lotka-Volterra systems are less stable than systems
with more realistic functional responses. It would be interesting to
see how the model \cite{laessig01} behaves if several trophic layers 
are allowed. We expect that it would be less stable and that the webs 
would be smaller than for model \cite{drossel01}.

The fact that the different evolutionary models give the
same overall structure of the food web, indicates that the shape of food webs 
results from a few general principles, and is not much affected by certain 
details of the system.  All models include competition between predators, 
and an efficiency of converting eaten prey into predator biomass which is 
smaller than 1. Furthermore, there are only few different external resources. 
Qualitatively, one can understand how these ingredients generate the 
familiar food web shapes: one can argue that very few external resources, 
combined with competition among the species feeding on these resources
and with predation allowing only efficient feeders to survive, 
leads to only a few basal species that feed on these resources. Since there
are more basal species than different external resources, there are more 
predator species that can coexist feeding on the basal species. The number 
of species thus increases initially with increasing trophic level, until 
the total biomass becomes so small that only a few top species can survive 
at the highest level. This explains the short length of food chains and 
the small proportion of basal and top species. The small number of links 
per species must result from the competition between predators, the strength 
of which can be tuned in all of the models. 

\section{Conclusions}
\label{conclusions}

Our aim in this review has been to survey most of the approaches 
that have been used to model food webs and to discuss real web data
in enough detail to provide a background to the model building. A
number of clear trends can be seen in both theory and experiment. In
the latter case there has been a realization in the last decade or so
that data collection needs to become a much more controlled and systematic 
affair. Much of the older data was collected as a by-product of other 
projects, and it has become clear that the database of collected webs contains
enough hidden biases to raise serious doubts about its usefulness. Recently 
food web data has been far more painstakingly collected, and has been done 
as part of projects specifically devoted to the investigation of food webs. 
Eventually the quality of the ensembles of collected webs will reflect this. 
The theory of food web structure has also made advances during this time. 
Much of the early work concerned webs with random, rather than evolved, 
structures, but there has been a tendency in later work towards building up 
a web either from a pool of species (assembly models) or, more recently, by 
creating webs through modification of the existing species (evolutionary 
models). In parallel with these developments, some suggestions for more 
realistic population dynamics for whole communities, and not just for a 
single predator-prey pair, have been put forward. In our opinion these last 
two developments, taken together, will form the basis for further progress 
in the theory of food webs.

We have seen that evolved food webs or webs generated through a series
of species invasions are generally more stable than randomly assembled
food webs, even if realistic food web shape is imposed.  Large and
complex randomly assembled food webs simply collapse under the
population dynamics and are very unlikely to be stable. Evolutionary
dynamics can chose the predator--prey links and the competition
structure such that stable webs are built step by step, starting from
a small initial web. We do not yet know in sufficient detail in what
respects the link strengths and link structure of evolved networks
differ from ad-hoc compositions. 

Furthermore, there is a need to investigate the effects of different
functional responses on the structure and stability of food web
models, rather than continuing to use the less satisfactory
Lotka-Volterra equations.  In section \ref{dynamicmodels} we drew
attention to the existence of theories of population dynamics with
realistic functional responses which can be applied to communities
with arbitrary structures.  The amount of computer time required to
construct model webs will not obviously be much greater than using
Lotka-Volterra equations, and there will probably be a saving due to
the increased stability of the models with more realistic functional
responses.  

To conclude, we hope that more realistic functional responses will be
used to investigate the structure and stability of assembled and
evolved webs in greater depth in the future, and that models will be
constructed that create webs by both immigration from a species pool
and by variation of the species within the community. We believe that
these, and other similar studies, will pave the way for a greatly
increased understanding of the structure and nature of food webs over
the next few years.

\section*{Acknowledgements}

We thank C. Quince for producing Figure 2. B.D. was supported by the Deutsche 
Forschungsgemeinschaft (DFG) under Contract No Dr300-2/1.

\end{document}